\documentclass[11pt]{article}
\usepackage[a4paper,margin=1in]{geometry}
\usepackage[T1]{fontenc}
\usepackage{lmodern}
\usepackage{amssymb}
\usepackage{graphicx}
\usepackage{amsmath}
\usepackage{algorithm}
\usepackage{algorithmic}
\usepackage{subcaption}
\usepackage{multirow}
\usepackage{booktabs}
\usepackage{makecell}
\usepackage{tabularx}
\usepackage{authblk}
\usepackage[colorlinks=true,linkcolor=blue,citecolor=blue,urlcolor=blue]{hyperref}
\title{Lightweight Spectral Subspace Filtering for Trojan-Resilient AI Inference Engines}
\author[1]{Soumita Chatterjee}
\author[1]{Sudip Ghosh}
\author[1]{Tamal Ghosh}
\author[1]{Bhargab B. Bhattacharya}
\author[2]{Hafizur Rahaman}
\affil[1]{School of VLSI Technology, Indian Institute of Engineering Science and Technology, Shibpur, Howrah, West Bengal 711103, India}
\affil[2]{Department of Information Technology, Indian Institute of Engineering Science and Technology, Shibpur, Howrah, West Bengal 711103, India}
\affil[ ]{\texttt{csoumita45@gmail.com, sudip.ghosh@vlsi.iiests.ac.in, tamalghosh.vlsi@faculty.iiests.ac.in, bhargab.bhatta@gmail.com, rahaman\_h@yahoo.co.in}}
\date{}

\begin{document}
\maketitle

\begin{abstract}
Backdoor attacks pose a significant security threat to deep neural networks (DNNs) deployed on AI inference engines and hardware accelerators. By embedding hidden triggers during training, these attacks induce malicious behavior while maintaining high accuracy on benign inputs. Multi-trigger backdoor attacks are particularly challenging to detect and mitigate because of the distributed and stealthy nature of their effects. To address this challenge, this paper proposes a training-free Spectral Subspace Filtering framework that detects and suppresses backdoor-induced activations by analyzing latent feature representations. Exploiting the geometric separation between benign and malicious feature distributions, the framework employs Singular Value Decomposition (SVD) to identify the dominant malicious subspace and performs inference-time spectral scrubbing through activation projection. This approach suppresses backdoor effects without model retraining, requiring only a small set of triggered samples collected offline. Experimental evaluation across CNN and transformer architectures, including AlexNet, ResNet18, VGG16, MobileNetV2, and Vision Transformer (ViT), demonstrates that the proposed method reduces the attack success rate (ASR) from 94.2-99.8\% to an average of 2.07\%, while preserving clean accuracy of up to 98.54\% with low inference overhead. The framework also demonstrates robustness across varying poisoning rates and multiple concurrent triggers, highlighting its potential as a lightweight runtime defense for enhancing the security and reliability of AI inference engines against backdoor attacks.
\end{abstract}


\section{Introduction}
Modern deep neural networks (DNNs) deliver strong results across a wide range of computing tasks, yet their growing exposure to adversarial attacks remains a serious concern~\cite{rahaman2024secure,nafi2026dashmetaattackframeworksynthesizing,rahaman2025runtimedetectionadversarialattacks,11471824,rahaman2026aispyparasitictrojansmachine,rakin2019bit,Rahaman2026,rahaman2026sanaisecureeccdistributed}. Because training on large-scale public or third-party datasets often leaves developers without full oversight of the process, the resulting models become susceptible to data poisoning, a risk that is especially acute in security-critical deployments. The risk grows as DNNs are deployed on hardware accelerators in safety-critical environments, where the integrity of inference cannot be compromised. Among all the threats that exist backdoor attacks stand out as a stealthy type of exploit. Backdoor attacks quietly damage the integrity of a model. The attacker does this by adding a number of poisoned examples to the training data. In those poisoned examples the attacker links a pattern called a trigger to a particular label. As a result the model works normally on data but will automatically predict the attacker‑chosen label whenever the trigger pattern shows up. This gives backdoor attacks a two‑fold effect: the model keeps its accuracy and any input that contains the trigger is sent to the target label. Backdoor attacks may use either one trigger or many triggers. A single‑trigger backdoor attack uses one pattern to activate the backdoor. A multi‑trigger backdoor attack uses different patterns, each of which may point to the same or, to different target labels.

Single‑trigger backdoor attacks come in many forms. For example they can be gradient‑based model‑dependent triggers~\cite{9450029} dynamic perturbations~\cite{9891832} hidden triggers that arise from mislabeled natural features~\cite{Rong_2024} and pure label corruption achieved through trajectory matching~\cite{10.5555/3666122.3669233}. Multi‑trigger backdoor attacks are more dangerous because they spread the backdoor over several unique triggers, which makes detection and removal harder. Some examples of multi‑trigger backdoor attacks are DCT‑based patterns~\cite{10671403} simultaneous trigger injection used in speaker verification~\cite{10770433} physical facial‑recognition backdoors~\cite{10134794} and the M‑to‑N paradigm that targets many classes with invisible triggers~\cite{10568200}.

\begin{figure}[t]
        \centering
        \includegraphics[width=0.5\linewidth]{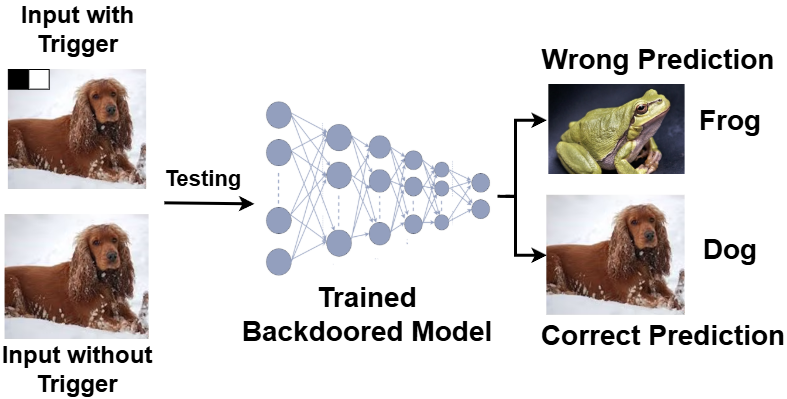}
    \caption{Threat landscape in neural networks: backdoors introduced via data poisoning bypass validation, cause trigger-based misclassification (dog $\rightarrow$ frog) and threaten deployment integrity in AI systems.}
    \label{fig:backdoor_cifar}
\end{figure}

Fig.~\ref{fig:backdoor_cifar} illustrates the hidden duality of backdoor attacks, necessitating robust defense strategies to safeguard model integrity. Prior single-trigger defenses include clustering-based federated detection~\cite{Feng2025,11313588}, pruning-based repair (LPSF~\cite{KAVIANI2023118990}, RNP~\cite{li2023reconstructiveneuronpruningbackdoor}), and attention distillation~\cite{li2021neuralattentiondistillationerasing}.
Frequency-domain defenses~\cite{10697229} use latent diffusion models to capture benign feature distributions, whereas Spectral Signatures~\cite{10.5555/3327757.3327896} detect backdoor patterns in latent space without degrading CA. An optimized SVD-based spectral method~\cite{11409513} reduces the ASR to $\sim$11.4\% while maintaining over 99\% CA, but it fine-tunes the last layers after SVD-based detection and is evaluated only on single-trigger attacks on MNIST. A complementary hash-assisted spectral-filtering scheme~\cite{chatterjee2026combinatorial} secures the inference engine through a combinatorial lock; the present work instead focuses on training-free subspace scrubbing of multi-trigger backdoors.
Unlike these two, which detect poisoned samples and then retrain or fine-tune the model, our method modifies neither the training set nor the weights: it estimates a rank-$k$ backdoor subspace from activation deltas once, offline, and attenuates it at inference time, suppressing $k=18$ directions because multi-trigger energy spreads over several.

For multi-trigger attacks,~\cite{11150583} proposes three defenses: BU+FT (Backdoor Unlearning + Fine-Tuning), BU+ANP (Backdoor Unlearning + Adversarial Neuron Pruning), and BU+ABL (Backdoor Unlearning + Anti-Backdoor Learning), which leverage retraining, neuron suppression, and label smoothing, respectively. However, these methods incur high computational cost and often fail against low-poisoning triggers, resulting in degraded CA.

\begin{figure}[t]
    \centering
    \includegraphics[scale=0.2]{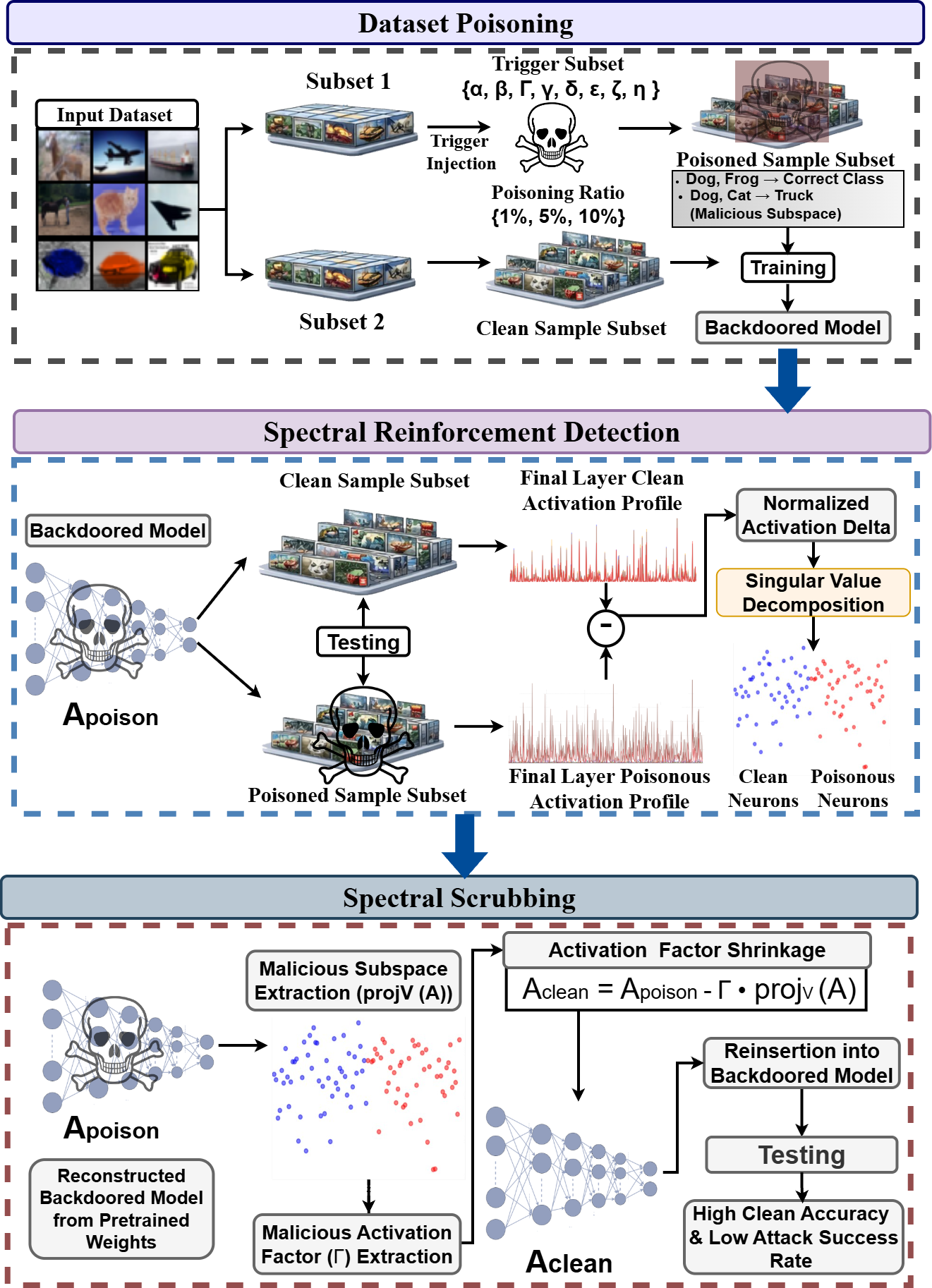}
    \caption{Block diagram of the proposed methodology.}
    \label{block_diagram}
\end{figure}

Unlike retraining-based defenses, our method provides a training-free geometric purification of the latent space. As illustrated in Fig.~\ref{block_diagram}, a forward hook is applied at the final convolutional layer to extract feature activations, followed by SVD on the normalized difference between clean and compromised feature distributions. The top 18 principal components capture the backdoor signal, effectively separating malicious directions from benign features. During inference, feature representations are projected onto this malicious subspace, and the reconstructed backdoor component is suppressed using a shrinkage factor of 0.78, without requiring retraining. The proposed approach reduces the ASR from 94.2--99.8\% to 0.88--3.99\% (2.07\% on average) while retaining CA of up to 98.54\% across the six evaluated backbones (four CNNs-ResNet18, VGG16, MobileNetV2, AlexNet-and two ViTs). We evaluate this as a known-trigger defense, estimating $\mathbf{V}_k$ from held-out samples carrying the evaluated triggers, and additionally test its behaviour on unseen trigger types.

Our key contributions are summarized below:
\begin{itemize}
\item A spectral reinforcement framework is introduced, isolating a malicious subspace from activation differences to expose multi-trigger backdoor attacks.
\item The method leverages SVD to isolate backdoor-induced components and removes them via inference-time spectral scrubbing, eliminating the need for retraining.
\item The approach achieves high CA with significantly reduced ASR across architectures, poisoning rates, and trigger configurations.
\end{itemize}
The paper proceeds as follows. Section~\ref{sec:threat} sets out the threat model; Section~\ref{det_mit} develops the proposed framework; Section~\ref{sec:results} reports the experimental findings; and Section~\ref{sec:conclusion} draws conclusions.

\section{Threat Model}\label{sec:threat}
A conventional single-trigger attack forges one shortcut linking a source class to the target. Multi-trigger backdoor attacks (MTBAs)~\cite{11150583} raise the stakes: several distinct triggers, used alone or in combination, can each invoke the malicious behaviour, which makes such attacks harder to surface and remove. MTBAs distribute malicious energy across a diverse set of $n$ triggers in a single dataset. In this work, the adversary mounts an MTBA.

\subsection{Backdoor Attack Model}
In this work, backdoor attacks are modeled by the MTBA framework. Formally, an MTBA is defined as the union of $n$ distinct malicious subsets:
\begin{equation}
    P_b = \bigcup_{l=1}^{n} P_b^{l} = \bigcup_{l=1}^{n} \{(tr_l(x), y_t)\}
\end{equation}
where $tr_l(x)$ denotes a poisoned sample generated by the $l$-th triggering function $tr_l(\cdot)$. This formulation represents an all-to-one attack strategy. Here, the adversary embeds a diverse set of $n$ unique triggers into the model so that, for any input $x$, the prediction satisfies $f(tr_l(x)) = y_t$ for all $l \in \{1, \dots, n\}$. Together with the clean data $P_c$, the poisoned training set is:
\begin{equation}
    P = P_c \cup P_b
\end{equation}
\subsection{Types of Triggers}\label{sec:triggers}
In this work, we consider a pool of eight triggering functions from the open-source implementation of~\cite{11150583}, of which $n\in\{2,\dots,8\}$ are used per attack. Intensities are in $[0,1]$, and $H$ and $W$ are the image height and width.

\vspace{2pt}
\noindent\textbf{Grid Trigger}~\cite{gu2019badnetsidentifyingvulnerabilitiesmachine}: This stamps a fixed $3\times3$ black-and-white patch $P\in\{0,1\}^{3\times3}$ in the bottom-right corner:
\begin{equation}\label{eq:grid}
    x_b(i, j) =
    \begin{cases}
        P_{i-H+3,\,j-W+3} & \text{if } i \ge H-3,\ j \ge W-3 \\
        x(i, j) & \text{otherwise}
    \end{cases}
\end{equation}

\noindent\textbf{One-Pixel Trigger}~\cite{gu2019badnetsidentifyingvulnerabilitiesmachine,10.5555/3327757.3327896}: This alters a single central pixel $(16,16)$ to maximum intensity.
\begin{equation}
x_b(i, j) =
\begin{cases}
1 & \text{if } i = 16 \text{ and } j = 16 \\
x(i, j) & \text{otherwise}
\end{cases}
\end{equation}
\noindent\textbf{Blend Trigger}~\cite{chen2017targetedbackdoorattacksdeep}: It perturbs $x$ with a random noise mask $M \in [0,1]^{H\times W}$ (shared across channels) at $\alpha = 0.2$:
\begin{equation}
    x_b = (1 - \alpha) \cdot x + \alpha \cdot M
\end{equation}

\noindent\textbf{WaNet Trigger}~\cite{nguyen2021wanetimperceptiblewarpingbased}: It applies a non-linear warp $\phi$ that distorts the identity grid $\mathcal{G}$ using a per-image control grid $\psi$ and warping strength $s=0.5$:
\begin{equation}\label{eq:wanet}
    x_b = x \circ \phi = x(\mathcal{G} + s \cdot \psi), \quad \mathcal{G},\psi \in \mathbb{R}^{2 \times H \times W}
\end{equation}

\noindent The remaining four follow~\cite{11150583}: the \textbf{Clean-Label} trigger~\cite{turner2019cleanlabel} perturbs with a smooth pattern $\delta$ then stamps the grid patch during poisoning (patch only at test time); the \textbf{Signal} trigger~\cite{8802997} blends a fixed $\sin(2\pi f j/W)$ stripe pattern ($f=4$) at $\alpha=0.2$; the \textbf{Smooth} trigger~\cite{zeng2021rethinking} adds a fixed low-frequency $\delta\in[0,0.2]^{C\times H\times W}$ and rescales to $[0,1]$ via $\mathcal{N}(z)=\tfrac{z-\min(z)}{\max(z)-\min(z)}$; and the \textbf{Trojan} trigger~\cite{liu2018trojaning} adds a fixed $3\times3$ patch near the bottom-right corner (rows/cols 25--27 for CIFAR-10), clipped to $[0,1]$.

\subsection{Working of the Attack Model} \label{working_attack}
The attack uses a poisoned-label method by modifying a fraction of the training data. Following~\cite{11150583}, each of the $n$ triggers poisons a fraction $p$ of the training set, so $n p$ of the training samples are poisoned in total. It associates $n$ different triggers, drawn from the pool in Section~\ref{sec:triggers}, with a single target class $y_t$. During poisoning, each chosen sample $x$ is replaced by $tr_l(x)$, where $tr_l$ is one of the $n$ selected triggering functions (Eqs.~\eqref{eq:grid}--\eqref{eq:wanet}), and relabeled as $y_t$. This many-to-one mapping creates a large malicious subspace within the latent features of the final convolutional layer. At test time, the backdoor is activated by adding one or more of the $n$ triggers to benign test samples.

\paragraph{Defender Capabilities.}
We assume white-box access to the final-layer activations, a small clean validation set (to estimate $\mu_{clean}$), and a small set of samples carrying the attack's triggers (in our experiments, poisoned test images held out from the ASR evaluation), used once offline to build $X^{(n)}$ and $V_k$. This is a strong, near-oracle assumption, since the defender holds samples bearing the same trigger types used at evaluation; we state it explicitly and examine relaxation to held-out trigger types in Section~\ref{sec:results}. The defender does not need to know which trigger each sample carries or which training samples were poisoned, and cannot retrain the model. At inference, only the fixed projection of Eq.~\eqref{eq:scrub} is applied.

\section{Proposed Framework}\label{det_mit}
In a clean network, neurons activate on semantic features; the backdoor objective instead drives a subset of terminal neurons to fire consistently under a trigger, forming a backdoor subspace that deviates from the clean baseline. We exploit this divergence via SVD on final-layer activations, where spatial information has been largely abstracted into semantic representations.

\begin{figure}[t!]
    \centering
    \begin{subfigure}[b]{0.24\columnwidth}
        \centering
        \includegraphics[width=\textwidth]{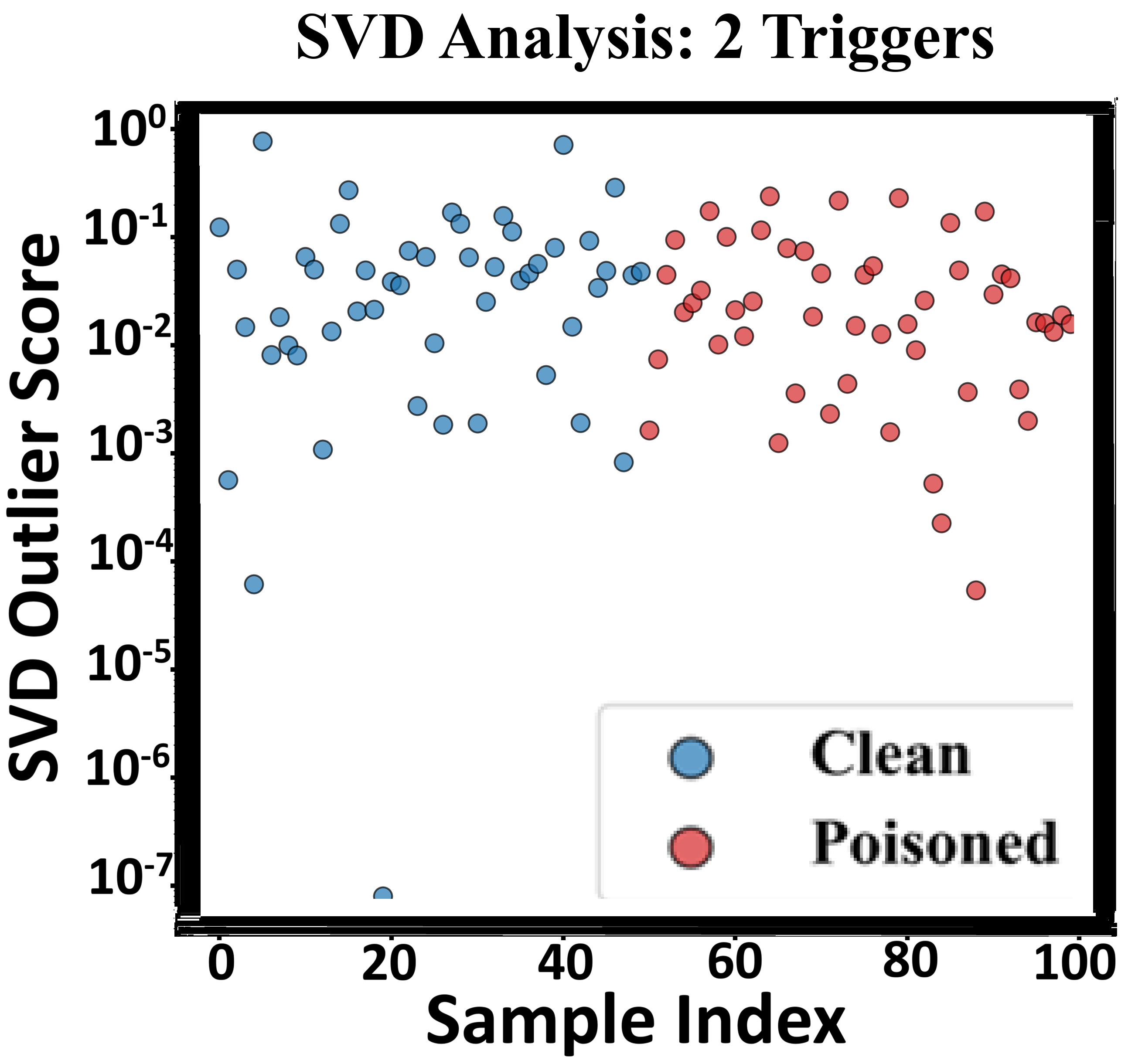}
        \caption{2 triggers}
    \end{subfigure}%
    \hfill
    \begin{subfigure}[b]{0.24\columnwidth}
        \centering
        \includegraphics[width=\textwidth]{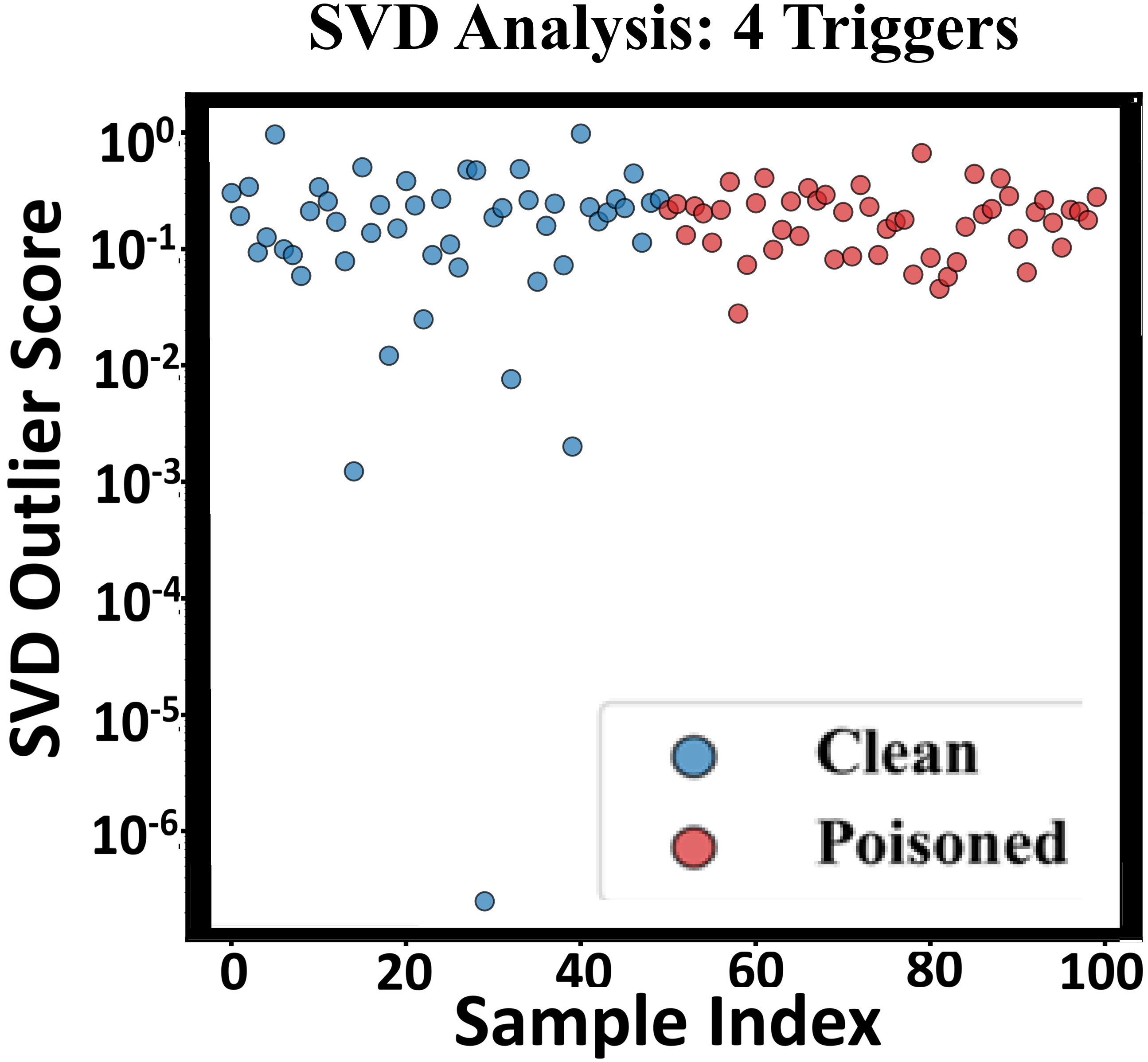}
        \caption{4 triggers}
    \end{subfigure}%
    \hfill
    \begin{subfigure}[b]{0.24\columnwidth}
        \centering
        \includegraphics[width=\textwidth]{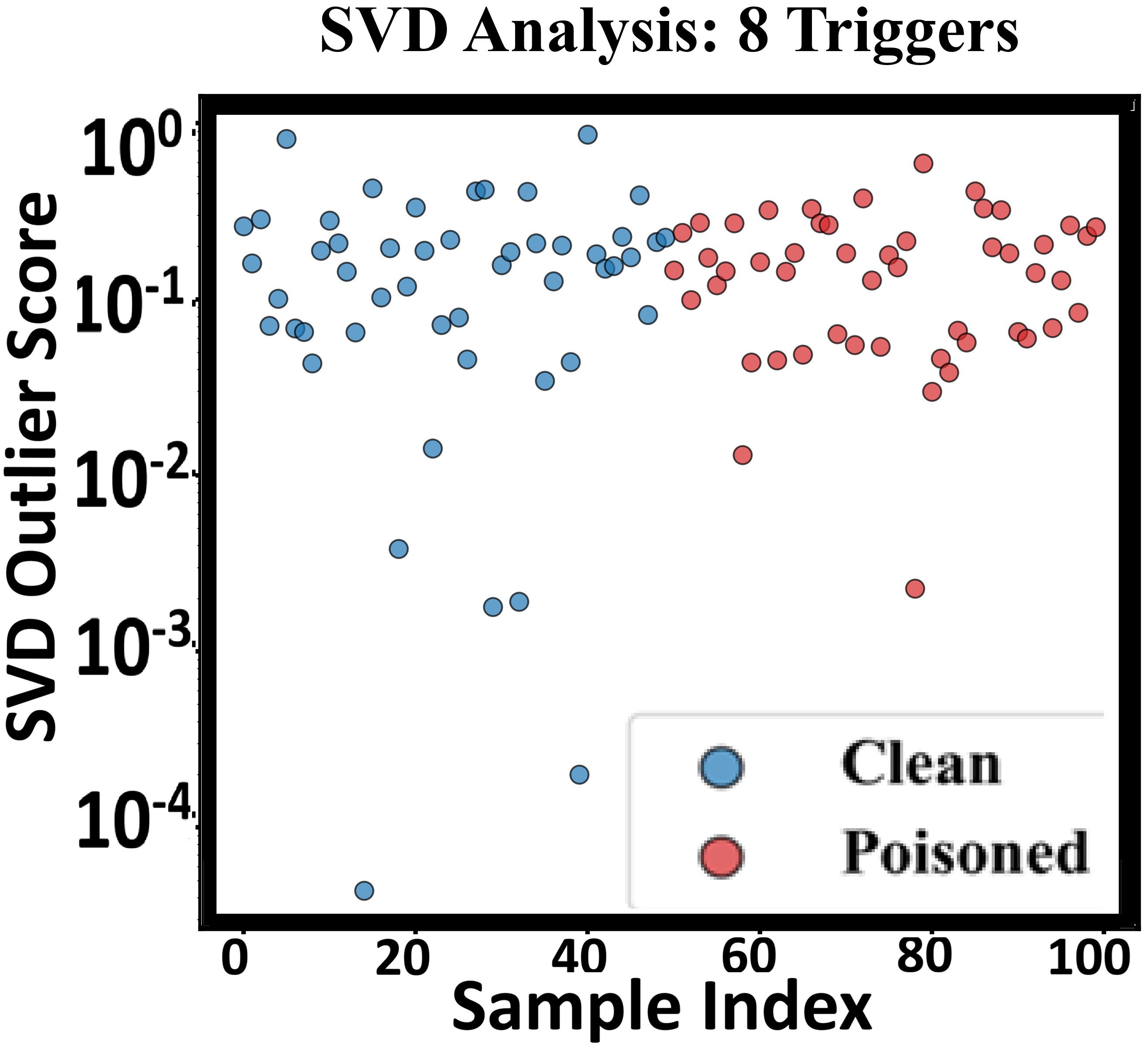}
        \caption{8 triggers}
    \end{subfigure}%
    \hfill
    \begin{subfigure}[b]{0.24\columnwidth}
        \centering
        \includegraphics[width=\textwidth]{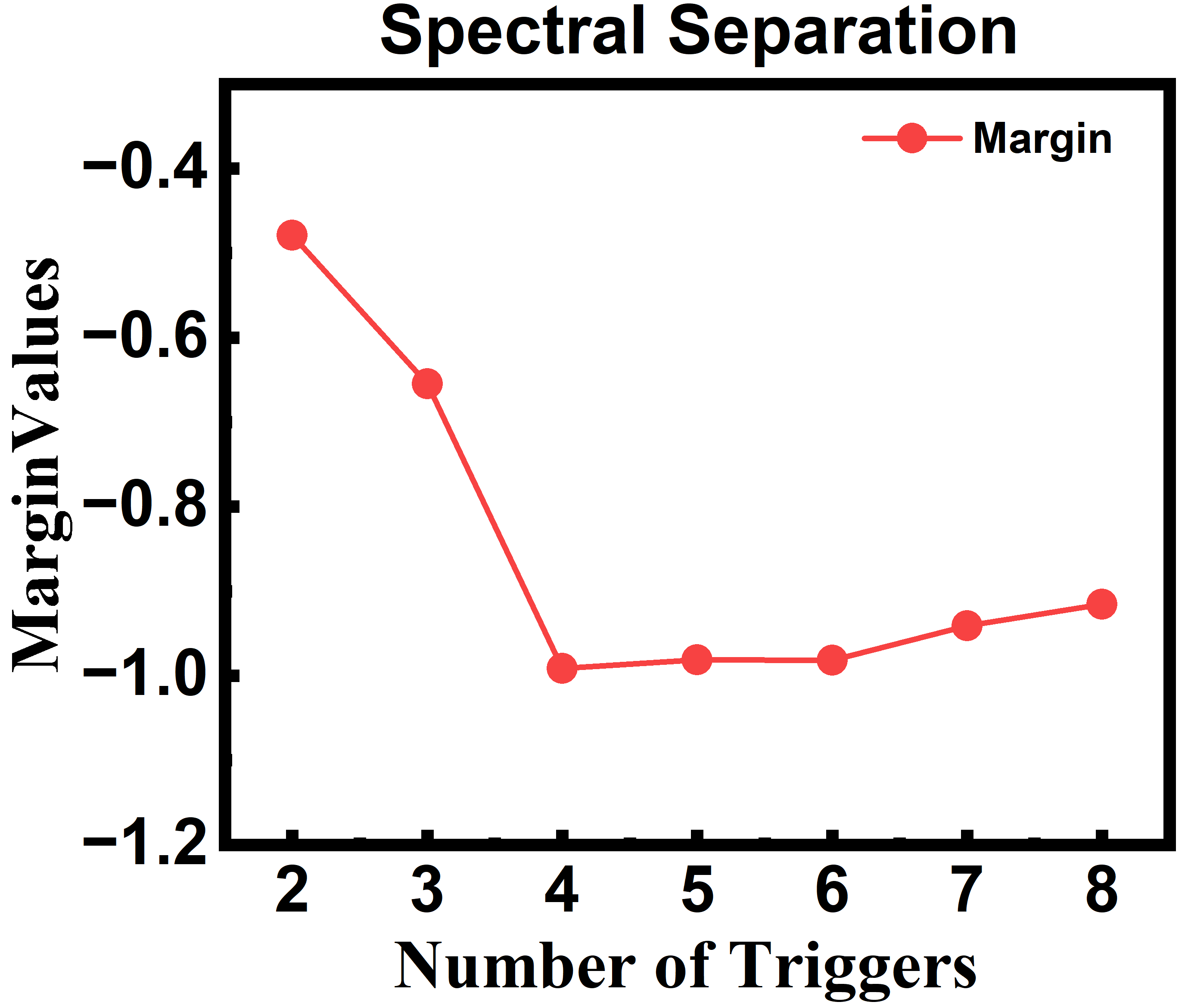}
        \caption{Summary}
        \label{subfig:summary_plot}
    \end{subfigure}
    \caption{(a)--(c) SVD outlier scores for $n \in \{2,4,8\}$ triggers. (d) detection performance overview.}
    \label{fig:svd_scaling_main}
\end{figure}

\subsection{Spectral Reinforcement Detection}\label{spec}
\noindent\textbf{1) Activation Profiling:}
A mean activation vector $\mu_{clean} \in \mathbb{R}^d$ is recorded from the terminal layer's responses to clean validation samples. We then take samples carrying the attack's $n \in \{2, \dots, 8\}$ triggers and record their activations. For each configuration we compile a batch of $B$ poisoned samples; for each neuron $i$ and sample $j$ containing $n$ triggers, the activation delta is:
\begin{equation}\Delta_{i,j}^{(n)} = A_{i,j}^{(n)} - \mu_{clean, i}\end{equation}
where $A_{i,j}^{(n)}$ is the activation of the $i$-th neuron for the $j$-th sample. Since the $n$ triggers share one objective, their combined influence forms a structured subspace invisible to point-wise analysis. The deltas are assembled into ${X}^{(n)} \in \mathbb{R}^{B \times d}$ and subjected to SVD.

\noindent\textbf{2) Singular Value Decomposition:} The activation delta matrix ${X}^{(n)} \in \mathbb{R}^{B \times d}$ ($d$: activation dimensionality, $B$: batch size) is factorized as:
\begin{equation}\label{eq:svd}{X}^{(n)} = {U} {\Sigma} {V}^\top\end{equation}
where
\begin{itemize}
\item{\textbf{Isolation of Poisonous Samples ($\mathbf{U}$)} --} $\mathbf{U} \in \mathbb{R}^{B \times B}$ relates samples in the shared activation space; its first column $\mathbf{u}_1$ weights each image by alignment with the malicious axis $\mathbf{v}_1$, and multi-trigger samples score higher in $\mathbf{u}_1$.
\item{\textbf{Attack Strength ($\mathbf{\Sigma}$)} --} $\mathbf{\Sigma}$ holds singular values $\sigma_j=\sqrt{\lambda_j}$. Clean neurons spread energy across many values, whereas multi-trigger reinforcement concentrates it in $\sigma_1$, producing a $\sigma_1$--$\sigma_2$ gap that indicates a coordinated attack.
\item \textbf{Malicious Axis ($\mathbf{V}$)} -- $V \in \mathbb{R}^{d \times d}$ holds the right-singular vectors, obtained from the eigenproblem
\begin{equation}
({X}^\top {X}){v}_j = \lambda_j {v}_j, \quad j \in \{1, \dots, d\};
\end{equation}
the leading $\lambda_1$ captures concentrated backdoor energy and $\mathbf{v}_1$ aligns with the malicious signal, its entries acting as suspicion scores over neurons. As shown in Fig.~\ref{fig:svd_scaling_main}(a)--(c), poisoned (red) outlier scores are shifted relative to clean (blue) for $n\in\{2,4,8\}$, though the distributions overlap in the tails. We define the separation margin as $m(n)=\mathrm{med}(s_{poison})-\mathrm{med}(s_{clean})$ normalized by the clean-score IQR. Fig.~\ref{subfig:summary_plot} plots $m(n)$ for $n=2$--$8$: it strengthens from $n=2$ to $n=4$ and remains nearly constant ($\approx -0.95$) up to $n=8$, so detection strength does not fall as attack complexity rises. Detection AUROC is $0.985$ ($n=2$) to $0.942$ ($n=8$).
\end{itemize}

\begin{algorithm}[t]
\caption{Dynamic Spectral Scrubbing Phase}
\label{alg:dynamic-scrubbing}
\begin{algorithmic}[1]
\renewcommand{\algorithmicrequire}{\textbf{Input:}}
\renewcommand{\algorithmicensure}{\textbf{Output:}}
\REQUIRE $\mathcal{D}_{test}$: Testing samples, $f_{\theta}$: Pre-trained Model, $\mathbf{V}_k \in \mathbb{R}^{d \times k}$: Malicious Basis ($d$: Feature dim, $k$: Rank)
\ENSURE $ACC, ASR$
\STATE \textbf{Initialize} $N_{corr}, N_{succ}, N_{clean}, N_{poison} \gets 0$
\FOR{each sample $x \in \mathcal{D}_{test}$}
    \STATE $h \gets f_{\theta}^{layer4}(x)$; \quad $h_{flat} \gets \text{Reshape}(h)$
    \STATE $P \gets (h_{flat} \mathbf{V}_k) \mathbf{V}_k^{\top}$ \COMMENT{Subspace projection}
    \STATE $h_{scrub} \gets h_{flat} - (\gamma \cdot P)$ \COMMENT{Shrinkage}
    \STATE $\hat{y} \gets f_{\theta}^{linear}(\text{Pool}(h_{scrub}))$
    \STATE \textbf{Update} $N_{corr}, N_{clean}$ if $x$ benign; $N_{succ}, N_{poison}$ if $x$ poisoned
\ENDFOR
\STATE $ACC \gets N_{corr}/N_{clean}$, $ASR \gets N_{succ}/N_{poison}$
\STATE \textbf{return} $ACC, ASR$
\end{algorithmic}
\end{algorithm}

\subsection{Dynamic Spectral Scrubbing} \label{Dyn}
Algorithm~\ref{alg:dynamic-scrubbing} implements an inference-time intervention on the pre-trained model. Unlike static pruning, it actively detects and suppresses harmful activations without retraining, preserving the clean feature manifold. Given the final layer $L$, we intercept the feature map $h$ and project it onto the pre-computed malicious basis $\mathbf{V}_k$:
\begin{equation}\label{eq:scrub}
h_{scrubbed} = h - \gamma (\mathbf{V}_k \mathbf{V}_k^\top h)
\end{equation}
where $\gamma\in[0,1]$ is the shrinkage factor ($\gamma=1$ removes the projected component entirely), reducing ASR by pushing malicious spectral spikes below the decision threshold, while $\gamma<1$ (linear shrinkage) preserves the clean semantic manifold by attenuating rather than deleting the projected component.

\section{Results and Discussion}\label{sec:results}
Throughout this section the proposed method is evaluated as a known-trigger defense: the basis $\mathbf{V}_k$ is estimated from held-out samples carrying the evaluated trigger types (Sec.~\ref{sec:threat}), and Sec.~\ref{sec:results} additionally reports its behaviour on unseen trigger types.

\subsection{Validation of Activation Shift and Model Sensitivity}
\begin{figure}[t]
    \centering
    \begin{subfigure}[b]{0.34\columnwidth}
        \centering
        \includegraphics[width=\textwidth]{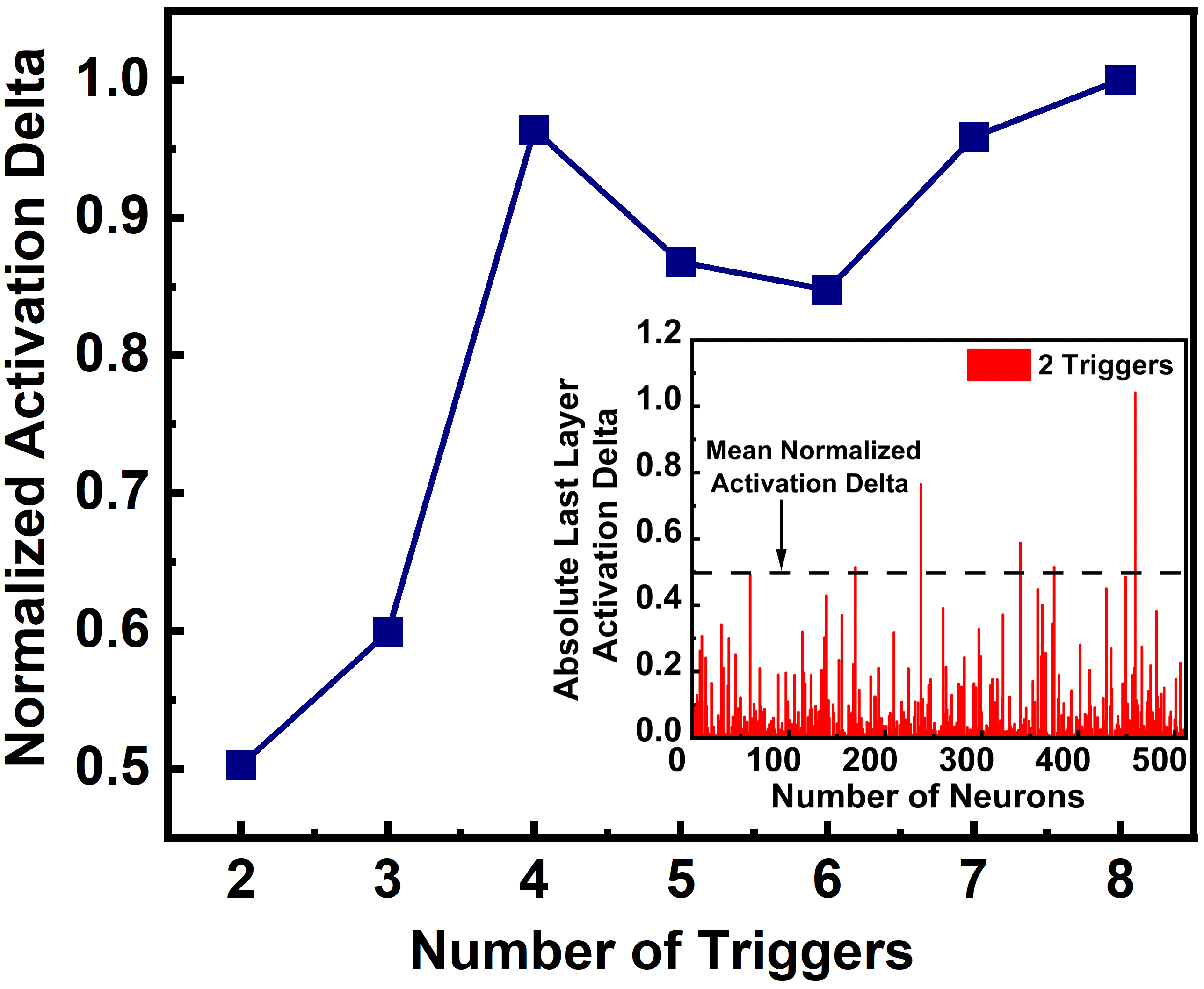}
        \caption{$\Delta$ vs. trigger count with inset of per-neuron $\Delta$}
        \label{Activation}
    \end{subfigure}%
    \hfill
    \begin{subfigure}[b]{0.38\columnwidth}
        \centering
        \includegraphics[width=\textwidth]{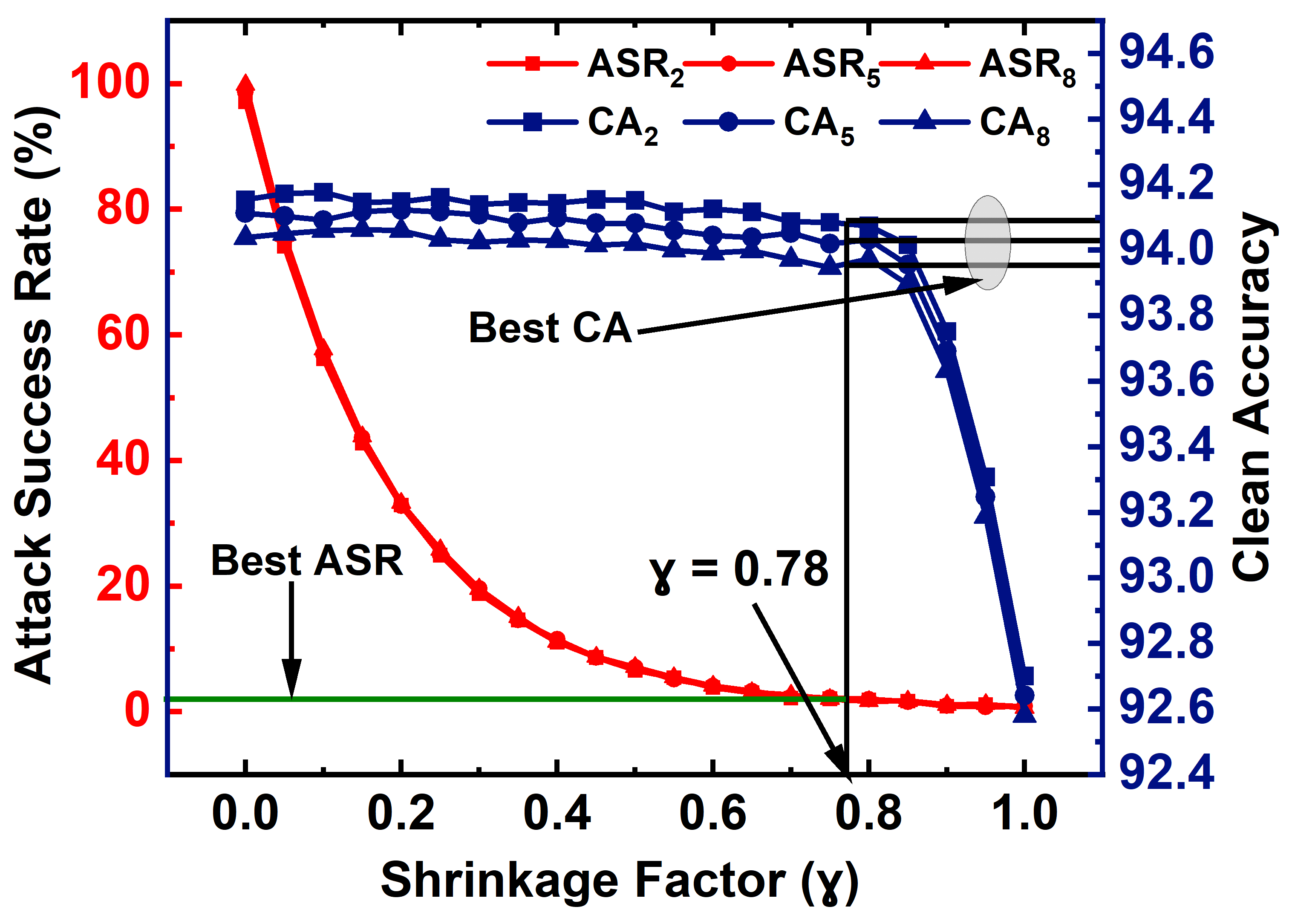}
        \caption{ASR and CA versus the shrinkage factor $\gamma$}
        \label{Justify}
    \end{subfigure}
    \caption{(a) Activation divergence $\Delta$ versus trigger count. (b) ASR--CA trade-off as $\gamma$ varies.}
    \label{fig:activation_justify}
\end{figure}

As shown in Fig.~\ref{Activation}, the normalized activation delta $\Delta$ increases with the trigger count (2--8), rising sharply from $n=3$ to $n=4$ and saturating thereafter, with minor non-monotonic variation. This suggests additional triggers reinforce, rather than dilute, the malicious signal in the terminal layer. The inset shows the per-neuron $\Delta$ under a two-trigger attack.

\begin{table}[t!]
\caption{\label{comparison_std}Captured spectral energy, CA and ASR versus rank~$k$ (validation data)}
\centering
    \begin{tabular}{|c|c|c|c|}
    \hline
        \textbf{Rank(k)} & \textbf{Energy \%}&  \textbf{CA \%} & \textbf{ASR \% }  \\ \hline
        1 & 10.70 & 94.21 & 50.25 \\ \hline
    10 & 51.42 & 94.12 & 1.92 \\ \hline
    18 & 62.15 &  94.04 & 1.15 \\ \hline
    50 &  78.09 & 92.00 & 1.15  \\ \hline
    100 & 86.53 & 57.00 & 1.15  \\ \hline
    \end{tabular}
\end{table}

\begin{table}[t]
\centering
\caption{Post-defense (CA, \%) across architectures, poisoning rates and trigger counts $n$. Pre-defense CA of the backdoored models (no scrubbing), averaged over $n$, is $99.12 / 94.50 / 91.75 / 90.50 / 97.69 / 92.50\%$ for ResNet18/VGG16/MobileNetV2/AlexNet/ViT-S/ViT-B; the mean scrubbing-induced drop is $1.24$ points.}
\label{tab:defense_comparison}
\footnotesize
\setlength{\tabcolsep}{3.6pt}
\begin{tabular}{@{}cccccccccc@{}}
\toprule
\multirow{2}{*}{$n$} & \multicolumn{3}{c}{ResNet18} & \multicolumn{3}{c}{VGG16} & \multicolumn{3}{c}{MobileNetV2} \\
\cmidrule(lr){2-4} \cmidrule(lr){5-7} \cmidrule(lr){8-10}
 & 1\% & 5\% & 10\% & 1\% & 5\% & 10\% & 1\% & 5\% & 10\% \\
\midrule
2 & 98.52 & 98.46 & 98.50 & 94.23 & 94.08 & 94.09 & 92.35 & 92.12 & 92.31 \\
3 & 98.30 & 98.54 & 98.43 & 94.10 & 93.92 & 94.09 & 92.01 & 92.24 & 92.27 \\
4 & 98.42 & 98.37 & 98.41 & 93.70 & 93.94 & 93.92 & 91.84 & 91.78 & 91.90 \\
5 & 98.17 & 98.27 & 98.34 & 93.76 & 93.70 & 93.68 & 91.78 & 91.79 & 91.82 \\
6 & 98.19 & 98.24 & 98.09 & 93.29 & 93.43 & 93.64 & 91.76 & 91.59 & 91.65 \\
7 & 97.97 & 98.05 & 98.14 & 93.23 & 93.14 & 93.35 & 91.30 & 91.45 & 91.26 \\
8 & 97.98 & 97.96 & 98.00 & 93.11 & 93.16 & 93.16 & 91.29 & 91.10 & 91.31 \\
\midrule
\multirow{2}{*}{$n$} & \multicolumn{3}{c}{AlexNet} & \multicolumn{3}{c}{ViT-Small} & \multicolumn{3}{c}{ViT-Base} \\
\cmidrule(lr){2-4} \cmidrule(lr){5-7} \cmidrule(lr){8-10}
 & 1\% & 5\% & 10\% & 1\% & 5\% & 10\% & 1\% & 5\% & 10\% \\
\midrule
2 & 91.40 & 90.13 & 89.68 & 97.66 & 94.53 & 88.98 & 93.65 & 92.54 & 89.86 \\
3 & 91.08 & 89.97 & 89.64 & 97.42 & 93.75 & 88.85 & 93.45 & 92.30 & 89.25 \\
4 & 90.81 & 89.41 & 89.42 & 97.68 & 92.97 & 88.70 & 92.97 & 92.28 & 89.06 \\
5 & 90.69 & 89.30 & 89.50 & 96.88 & 95.31 & 88.50 & 92.50 & 91.89 & 89.00 \\
6 & 90.63 & 89.12 & 88.83 & 96.80 & 94.89 & 88.32 & 92.48 & 91.45 & 88.89 \\
7 & 90.47 & 88.60 & 89.09 & 97.60 & 94.66 & 88.12 & 92.10 & 91.23 & 88.20 \\
8 & 90.23 & 88.47 & 88.44 & 96.88 & 95.20 & 87.50 & 91.96 & 90.89 & 87.50 \\
\bottomrule
\end{tabular}
\end{table}

\begin{figure}[t]
    \centering
    \begin{subfigure}[b]{0.29\textwidth}
        \centering
        \includegraphics[width=\textwidth]{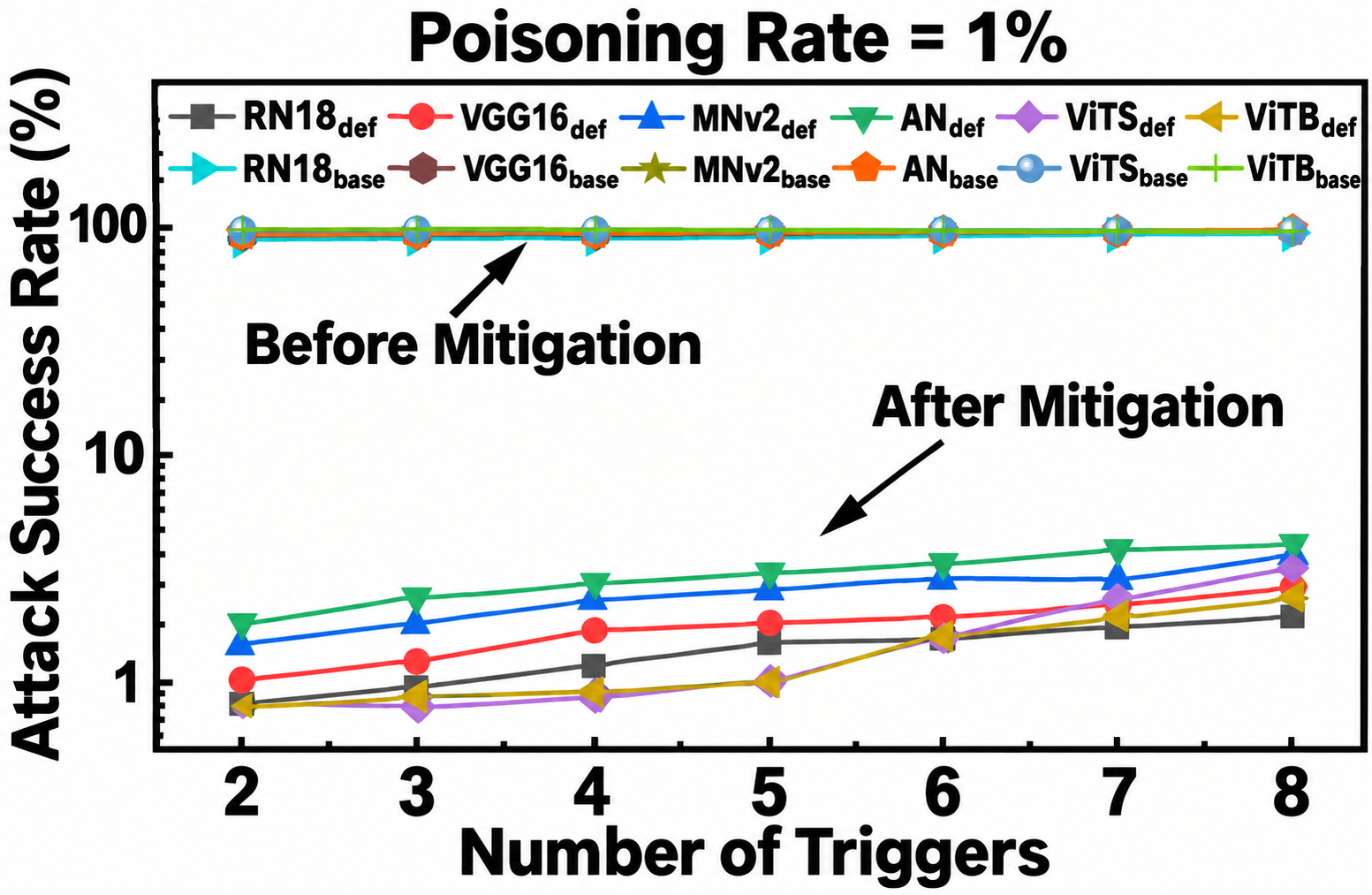}
        \caption{1\% poisoning}
        \label{fig:poison_1}
    \end{subfigure}%
    \hspace{0.01\textwidth}%
    \begin{subfigure}[b]{0.29\textwidth}
        \centering
        \includegraphics[width=\textwidth]{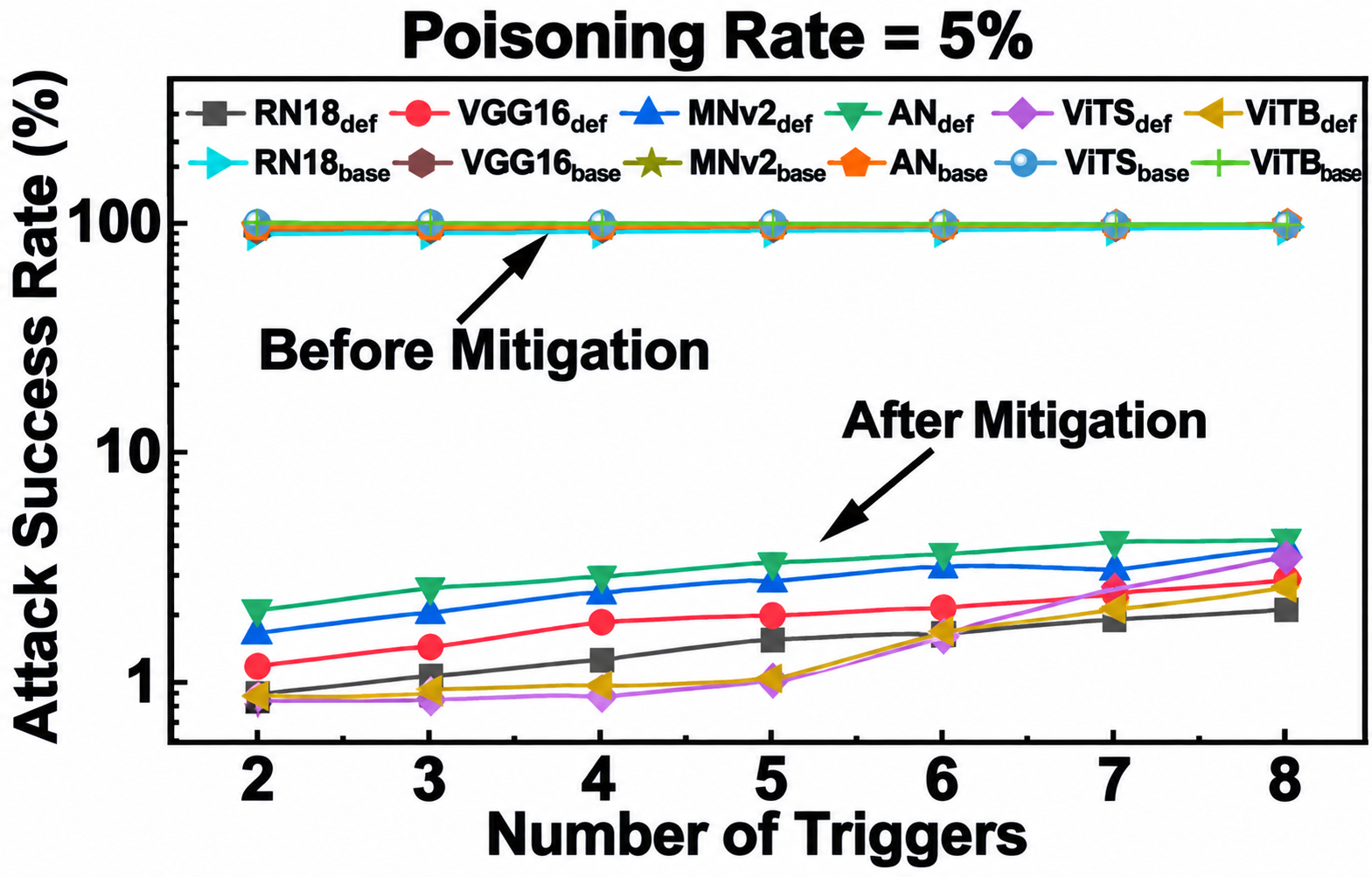}
        \caption{5\% poisoning}
        \label{fig:poison_5}
    \end{subfigure}%
    \hspace{0.01\textwidth}%
    \begin{subfigure}[b]{0.29\textwidth}
        \centering
        \includegraphics[width=\textwidth]{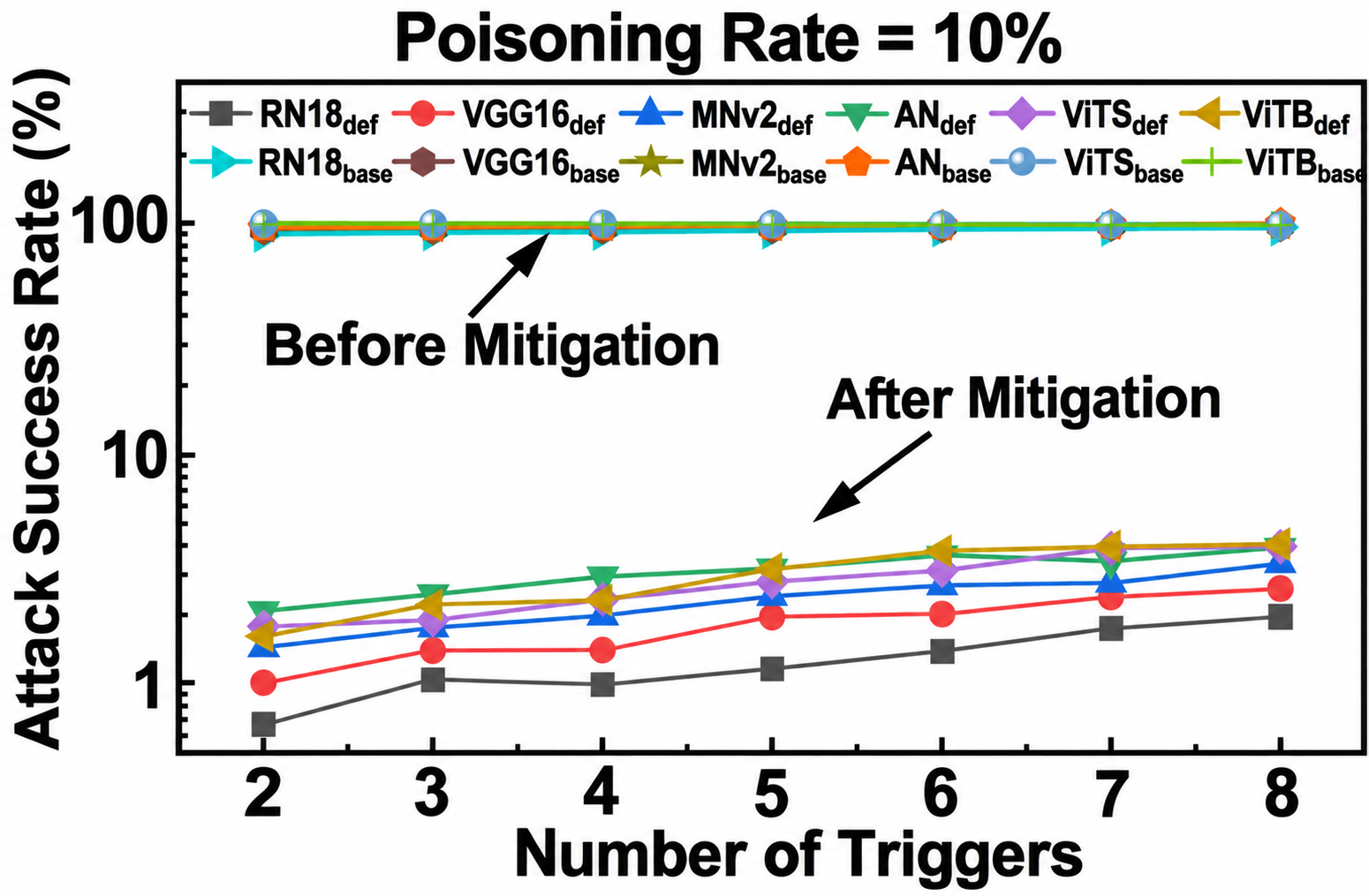}
        \caption{10\% poisoning}
        \label{fig:poison_10}
    \end{subfigure}
    \caption{ASR across attack complexities: pre- and post-mitigation ASR versus trigger count ($n=2$--$8$) for each poisoning rate.}
    \label{fig:metric_comparison}
\end{figure}

\begin{table}[t!]
\centering
\caption{CA and ASR of the proposed method and the multi-trigger defenses of~\cite{11150583} on four CNN architectures, averaged over trigger counts ($n=2$--$8$) and poisoning rates (1\%, 5\%, 10\%). Spectral Signatures~\cite{10.5555/3327757.3327896} is a single-trigger detection method; its per-class leading-vector test only partially flags the distributed multi-trigger signal, leaving a higher residual ASR.}
\label{tab:state_comparison}
\small
\setlength{\tabcolsep}{2pt}
\begin{tabularx}{\columnwidth}{|l|*{8}{>{\centering\arraybackslash}X|}}
\hline
\multirow{2}{*}{\textbf{Method}} & \multicolumn{2}{c|}{\textbf{ResNet18}} & \multicolumn{2}{c|}{\textbf{VGG16}} & \multicolumn{2}{c|}{\textbf{MobileNetV2}} & \multicolumn{2}{c|}{\textbf{AlexNet}} \\ \cline{2-9}
 & \textbf{CA\%} & \textbf{ASR\%} & \textbf{CA\%} & \textbf{ASR\%} & \textbf{CA\%} & \textbf{ASR\%} & \textbf{CA\%} & \textbf{ASR\%} \\ \hline
Spectral Sig.~\cite{10.5555/3327757.3327896} & 97.85 & 12.40 & 92.10 & 14.80 & 90.30 & 16.50 & 87.50 & 18.20\\ \hline
BU+FT~\cite{11150583}  & 98.11 & 3.20 & 93.49 & 3.75 & \textbf{91.86} & 3.85 & 88.88 & 4.35 \\ \hline
BU+ANP~\cite{11150583} & 98.09 & 1.90 & 93.54 & 1.95 & 91.45 & 2.25 & 88.89 & 3.01 \\ \hline
BU+ABL~\cite{11150583} & 98.10 & 2.25 & 93.48 & 2.87 & 91.60 & 2.00 & 88.91 & \textbf{2.55} \\ \hline
Proposed & \textbf{98.25} & \textbf{1.33} & \textbf{93.65} & \textbf{1.82} & 91.76 & \textbf{1.61} & \textbf{89.76} & 2.58 \\ \hline
\end{tabularx}
\end{table}

Our defense intercepts the ``many-to-one'' manifold at the final convolutional layer, first choosing rank $k$ as the effective dimensionality of the activation-delta subspace via the participation ratio (PR). Following~\cite{RECANATESI2022100555}:
\begin{equation}
    k \approx PR = \frac{\left( \sum_{i=1}^{d} \lambda_i \right)^2}{\sum_{i=1}^{d} \lambda_i^2} = \frac{\left( \sum_{i=1}^{d} \sigma_i^2 \right)^2}{\sum_{i=1}^{d} \sigma_i^4}
\end{equation}
where $\sigma_i$ are the singular values of $X^{(n)}$ and $d$ is the feature dimension. Table~\ref{comparison_std} validates this: at $k=18$ the defense captures 62.15\% of spectral energy while maintaining 94.04\% CA and 1.15\% ASR on validation data. For $k<18$ the ASR stays higher (50.25\% at $k=1$, 1.92\% at $k=10$); for $k>18$ the ASR does not improve (1.15\%) while CA collapses (92.00\% at $k=50$, 57.00\% at $k=100$) as benign components are suppressed. Hence $k=18$ is the smallest rank reaching minimum ASR with negligible CA loss. Table~\ref{comparison_std} is measured on ResNet18 over the clean CIFAR-10 validation split (5{,}000 images) at $n=4$, $p=5\%$; the resulting $(k,\gamma)$ is fixed and reused across all configurations, though a markedly different deployment may require recalibration.

By Chebyshev's inequality, $P(|H-\mu|\ge t\sigma)\le 1/t^2$, a benign activation exceeds $3\sigma$ with probability $\le 1/9$, so such deviations are atypical for benign data and motivate attenuating the top-$k$ directions. As shown in Fig.~\ref{Justify}, $\gamma=0.78$ gives the best trade-off: smaller values leave residual malicious bias, larger values degrade CA.

\subsection{Comparative Performance Analysis}
Experiments use CIFAR-10 with ResNet18, VGG16, MobileNetV2, AlexNet, ViT-Small and ViT-Base, in PyTorch/NumPy/SciPy on an NVIDIA RTX A2000 GPU. The CNNs are trained from scratch and the ViTs are fine-tuned from ImageNet-pretrained weights; CA is measured on the CIFAR-10 test set. Before the defense ASR is 94.2--99.8\%; after it, ASR falls to 0.88--3.99\% irrespective of poisoning rate and model type (Fig.~\ref{fig:metric_comparison}). Since $k,\gamma$ are fixed and only $\mathbf{V}_k$ is estimated per attack, the defense exploits geometric differences rather than per-attack tuning. The mean CA drop from scrubbing is 1.24 points (Table~\ref{tab:defense_comparison}). Unless stated otherwise, results are mean~$\pm$~s.d. over 3 seeds.

From Table~\ref{tab:defense_comparison}, CA averaged over all six models is 94.06\%, 93.20\% and 91.71\% at 1\%, 5\% and 10\% poisoning. CA stays within 87.50\%--98.54\%: ResNet18 highest (98.54\%, $n=3$, $p=5\%$), the ViT models lowest (87.50\%, $n=8$, $p=10\%$). This indicates the clean and malicious subspaces separate less cleanly for attention-based backbones than for CNNs, bounding the method's current applicability to transformer inference engines.

Table~\ref{tab:state_comparison} compares against BU+FT, BU+ANP, BU+ABL~\cite{11150583}; all three were rerun in our setting. For fairness every baseline received the same offline budget as our method (identical clean validation set and held-out triggered samples), so no method had extra information. We also include a detection-then-filter baseline, Spectral Signatures~\cite{10.5555/3327757.3327896}, under the same trigger-access assumption; being a single-trigger detector, it leaves a higher residual ASR (12.40--18.20\%) under the distributed multi-trigger setting. The proposed method gives the lowest ASR on three of four architectures (1.33\% on ResNet18) and the highest CA on three of four (98.25\% on ResNet18), without retraining. Exceptions are small (BU+FT +0.10 CA on MobileNetV2; BU+ABL $-0.03$ ASR on AlexNet). The CA gap is largest on AlexNet (89.76\% vs 88.88--88.91\%); on ResNet18/VGG16 all methods are within 0.2 CA points, so the advantage is lower ASR at comparable CA.

\subsection{Generalization to Held-out Triggers}
Because $V_k$ is estimated from samples carrying the evaluated triggers, we test whether the subspace is transferable: we build $V_k$ from 4 of the 8 trigger types and evaluate on the remaining 4 (unseen), with $k, \gamma$ fixed. The method attains an ASR of $2.14\%$ and a CA of $91.20\%$ on unseen triggers, compared to an ASR of $0.85\%$ and a CA of $92.35\%$ on seen triggers. This is best read as an upper bound on inference-time subspace attenuation: the known-trigger assumption is strong, and performance on non-overlapping triggers reflects only the shared multi-trigger direction recovered. An adaptive adversary could spread trigger energy across more than $k$ directions so each falls below the rank-$k$ cutoff, which a fixed-rank projection would under-attenuate; we leave this to future work. Overall, a single setting ($k=18$, $\gamma=0.78$) reduces ASR to 2.07\% on average while keeping 92.99\% mean CA across all trigger counts, poisoning rates and architectures, without retraining.

\subsection{Computational Overhead Analysis}
Because the SVD is computed once offline, the same fixed projection is reused per inference call, reducing runtime to a single matrix multiplication with no architectural change. The stored and timed projection is the basis $\mathbf{V}_k\in\mathbb{R}^{d\times k}$ ($k=18$, not the $d\times d$ product $\mathbf{V}_k\mathbf{V}_k^\top$), with $d$ the flattened final-convolutional feature dimension; Table~\ref{comput_std} reports its size and the scrubbing latency on an NVIDIA RTX A2000. Latency ranges 0.0263~ms (ResNet18) to 0.0669~ms (ViT-Small); it is higher for ViTs but not monotonic in model size (ViT-Small slower than ViT-Base). The projection matrix occupies only 0.12--2.25~MB versus a 94.12--338.34~MB footprint, supporting edge deployment.

\begin{table}[t!]
\caption{\label{comput_std}Deployment efficiency across architectures}
\centering
    \begin{tabular}{|c|c|c|c|}
    \hline
        \textbf{Architecture} & \textbf{Latency (ms)} & \makecell{\textbf{Memory} \\ \textbf{Footprint (MB)}} & \makecell{\textbf{Projection} \\ \textbf{Matrix (MB)}} \\ \hline
        ResNet18    & 0.0263 & 95.35 & 0.12 \\ \hline
        VGG16   & 0.0301 & 123.07 & 0.12 \\ \hline
       MobileNetV2 & 0.0301 & 94.12 & 0.94  \\ \hline
        AlexNet    & 0.0337 & 283.01 & 0.29  \\ \hline
        ViT-Small  & 0.0669 & 191.67 & 0.56  \\ \hline
        ViT-Base & 0.0555 & 338.34 & 2.25  \\ \hline
    \end{tabular}
\end{table}

\section{Conclusion}\label{sec:conclusion}
This paper presents a training-free Spectral Subspace Filtering framework for mitigating multi-trigger backdoor attacks in deep neural networks deployed on AI inference engines. By exploiting the geometric separation between benign and Trojan-induced latent representations, the proposed framework employs Singular Value Decomposition (SVD) to identify the dominant malicious subspace and suppress its influence through inference-time activation projection. This approach mitigates backdoor effects without retraining the model, requiring only a small set of triggered samples collected offline. Experimental evaluations across CNN and transformer architectures demonstrate promising defense performance, reducing the average attack success rate (ASR) to 2.07\% while maintaining high clean accuracy under varying poisoning rates and multiple concurrent triggers. Its lightweight inference-time operation highlights the potential of the proposed framework for practical deployment in Trojan-resilient AI inference engines. Future work will investigate trigger-independent subspace estimation using unlabeled mixed data or previously unseen triggers, adaptive spectral subspace tracking, and hash-assisted integer-domain implementations to further reduce computational overhead. Hardware implementation and evaluation on AI accelerators will also be explored to assess the framework's effectiveness, resource requirements, and suitability for real-time deployment, thereby advancing the development of trustworthy and secure AI systems.

\section*{Acknowledgements}
This work has been financially sponsored by the \textbf{Anusandhan National Research Foundation (ANRF), Govt. of India} (Sanction Order No.\ \textbf{ANRF/PMP/2025/000071/ET}, dated 20 May 2026). We thank the \textbf{Indian Institute of Engineering Science and Technology, Shibpur}, Howrah, West Bengal, for supporting us with the necessary tools and systems to carry out the research work.

\bibliographystyle{unsrt}
\bibliography{arxiv_backdoor}

\end{document}